\documentclass{elsart}
\usepackage{epsfig}
\usepackage{lineno}

\begin{document}
\runauthor{Marco Battaglia}
\begin{frontmatter}
\title{Characterisation of\\ a Thin Fully Depleted SOI Pixel Sensor\\ with High Momentum Charged Particles}
\author[LBNL,UCSC]{Marco Battaglia,}
\author[INFN]{Dario Bisello,}
\author[LBNL]{Devis Contarato,}
\author[LBNL]{Peter Denes,}
\author[INFN]{Piero Giubilato,}
\author[INFN]{Serena Mattiazzo,}
\author[INFN]{Devis Pantano}
\address[LBNL]{Lawrence Berkeley National Laboratory, 
Berkeley, CA 94720, USA}
\address[UCSC]{Santa Cruz Institute of Particle Physics, University of California 
at Santa Cruz, CA 95064, USA}
\address[INFN]{Dipartimento di Fisica, Universit\'a di Padova and INFN,
Sezione di Padova, I-35131 Padova, Italy}

\begin{abstract}
This paper presents the results of the characterisation of a thin, fully depleted pixel sensor manufactured in SOI technology 
on high-resistivity substrate with high momentum charged particles. The sensor is thinned to 70~$\mu$m  and a thin phosphor 
layer contact is implanted on the back-plane. Its response is compared to that of thick sensors of same design in terms of 
signal and noise, detection efficiency and single point resolution based on data collected with 300 GeV pions at the CERN SPS. 
We observe that the charge collected and the signal-to-noise ratio scale according to the estimated thickness 
of the sensitive volume and the efficiency and single point resolution of the thinned chip are comparable to those measured 
for the thick sensors.
\end{abstract}
\end{frontmatter}

\typeout{SET RUN AUTHOR to \@runauthor}

%\linenumbers

\section{Introduction}

Monolithic pixel sensors on high resistivity Si are a very attractive option for high resolution particle tracking.
Silicon on Insulator (SOI) is one of the leading technologies for manufacturing these devices with the possibility 
of integrating advanced data processing capabilities. With the mitigation of the back-gating effect by implanting 
a buried $p$-well (BPW) beneath the buried oxide (BOX)~\cite{soi2}, SOI pixel sensor prototypes have demonstrated  
high detection efficiency and micron-size single point resolution~\cite{beam2010-soi}. Because of the need to minimise 
multiple scattering in precision vertex tracking at future colliders, the total thickness of sensor ladders should be 
$\le$ 100~$\mu$m of Si-equivalent while retaining high S/N and detection efficiency~\cite{Battaglia:2010ce}. 
In this paper, we present the first characterisation of a thin, fully depleted SOI pixel sensor, with a nominal thickness 
of 70~$\mu$m, using high momentum charged hadrons. The performance of the thinned SOI sensor is compared to that of thick 
sensors of the same design.

\section{Thin SOI Sensor, experimental setup and data analysis}

The prototype chip used in this study is the ``SOImager-2'' sensor designed at Berkeley and manufactured by Lapis 
Semiconductor Co.\, Ltd.\ (formerly OKI Semiconductor) in 0.2~$\mu$m SOI technology on $n$-type SOI wafers with a nominal 
resistivity of the handle wafer of 700~$\Omega\cdot$cm. Its sensitive area is a 3.5$\times$3.5~mm$^2$ matrix of 256$\times$256 
pixels arrayed on a 13.75~$\mu$m pitch and read out through four parallel arrays of 64 columns each~\cite{beam2010-soi}. 
The chip implements pixel cells of different design with various combinations of floating $p$-type guard rings and 
BPW, of which we employ here two sectors, where the pixel electronics is protected by a BPW to avoid the back-gating effect. 
A grid of $p$-type guard-rings surrounds the I/O electronics at the chip periphery, while an external guard-ring surrounds 
the entire sensor design. This sensor has already been successfully tested with high momentum particles at the CERN SPS in 
2010~\cite{beam2010-soi}. The sensor under test is back-thinned using a commercial grinding technique~\cite{aptek}  which 
has been already successfully employed for back-thinning CMOS Active Pixel Sensors~\cite{backthin}.  
The sensor thinning and post-processing are presented in detail in another paper, where we discuss the chip response to 
X-rays~\cite{xray-soi}. The thickness of the thinned chip is measured to be to ($73 \pm 2$)~$\mu$m.
After thinning, a thin entrance window is created on the back-plane by implanting a thin phosphor layer contact and then 
the chip is annealed. The thickness of the P implant is measured using spreading resistance analysis (SRA) on a chip which 
indicates that the P implant extends to a depth of $\simeq$0.4~$\mu$m from the back-plane surface, with the highest 
concentration in the first 0.2~$\mu$m.
We estimate the thickness of the high resistivity handle wafer to be (64$\pm$3)~$\mu$m from the measured device thickness and 
foundry data, where the uncertainty is from the sensor and the back-plane implant thickness. These sensors can be fully 
depleted at voltage values of $\sim$40~V, well below the measured breakdown voltage of $\sim$130~V, which instead prevents 
full depletion at their 260~$\mu$m full thickness, as provided by the foundry. The pixels used in this study have a cell design 
with no $p$-type guard-ring and have the BPW connected to the pixel diode. The charge-to-voltage conversion of thinned and 
processed sensors is measured to be (31.3$\pm$0.4)~$e^-$ ADC count$^{-1}$ at 50~V, from their response to X-rays of various 
energies~\cite{xray-soi}.

In order to study the response of the thinned and back-processed sensor to minimum ionising particles and compare it to 
that of the unprocessed chips, three layers of SOI sensors have been installed on the CERN SPS H4 beam-line in September 2011.
The thinned SOI chip has been placed upstream from a doublet made of the same SOI chips with full thickness. The doublet was 
already used in the 2010 beam test data taking~\cite{beam2010-soi}. The setup has been exposed to a 300~GeV $\pi^-$ beam. 
The data acquisition system consists of an analog board pigtailed to a commercial FPGA development board, used as control 
unit, as in previous tests~\cite{Battaglia:2009daq}. The analog outputs from the three chips are connected to 
independent analog differential inputs, each feeding a 100~MS/s 14-bit ADC. The control board is equipped with a Xilinx Virtex-5 
FPGA supplying the clocks and the slow control signals driving the sensor chips and routing the digitised data to a high speed 
FIFO. Data are formatted and transferred to the DAQ computer via a USB-2.0 link at a rate of 25~Mbytes/s. Measurements are performed 
with the chip clocked at 12.5~MHz, corresponding to an 80~ns read-out time per pixel. The noise of the preamplifier stage and the 
read-out chain is 1.8~ADC counts. Data sparsification is performed on-line in the DAQ PC using custom 
{\tt Root}-based~\cite{Brun:1997pa} software. 
Sensors are scanned for seed pixels with signal exceeding a preset threshold in noise units. For each seed, the 7$\times$7 pixel 
matrix centred around the seed position is selected and stored on file. The pixel pedestal and noise values are updated at the end 
of each SPS spill in order to follow possible drifts in their baselines. Data are stored in {\tt Root} format and subsequently 
converted into {\tt lcio} format~\cite{Gaede:2005zz} for offline analysis. The data analysis is based on a set of custom processors 
in the {\tt Marlin} reconstruction framework~\cite{Gaede:2006pj} to perform cluster reconstruction and analysis, pattern recognition 
and track fitting~\cite{Battaglia:2008nj}. In the offline analysis of the sparsified data, clusters are reconstructed 
applying a double threshold method on the matrix of pixels selected around a candidate cluster seed. Clusters are requested to have 
a seed pixel with a signal-to-noise ratio, S/N, of at least 8.0 and the neighbouring pixels with a S/N in excess of 5.0. Clusters 
consisting of a single pixel are discarded. The hit position is calculated from the centre of gravity of the pulse height of 
pixels associated to a cluster. 
Particle tracks are reconstructed using a straight line model, since the multiple scattering can be safely neglected at 300~GeV. 
The detector planes are surveyed before data taking and these positions are used as the starting point of the offline alignment with 
particle tracks. Tracks are reconstructed from the space points obtained in the two layers of the doublet and extrapolated upstream 
onto the thin sensor. Given the low particle density and the relatively high read-out speed, there are on average only 1.13~hits/layer 
in non-empty events. This greatly simplifies the pattern recognition. For associating a hit to the track extrapolation a 
50$~\mu$m-wide window is used. The slope of candidate tracks is requested to be smaller than 3$\times$10$^{-3}$  in both 
coordinates, to remove particles originating from interactions and low-momentum secondaries. When using all the three layers to 
reconstruct the track we also require the fit $\chi^2$ to be below 5. When we study the efficiency and the single point resolution 
we use only the pair of hits on the doublet to define the track. The setup is simulated using the {\tt Geant-4} simulation 
toolkit~\cite{g4}. Charge collection in the depleted Si thickness and signal generation on the pixels is simulated using 
{\tt PixelSim}, a dedicated digitisation module~\cite{Battaglia:2007eu}, where a user parameter controls the charge diffusion 
on the pixels. The simulation is calibrated by inputting the measured single pixel noise and its spread and by adjusting the charge 
spread parameter to reproduce the observed pixel multiplicity in clusters associated to reconstructed tracks. Simulated digitised 
hits are then processed and reconstructed using the same {\tt Marlin} processors as the data. 

\section{Results}

\subsection{Leakage current and depletion thickness}

\begin{figure}[hb!]
\begin{center}
\epsfig{file=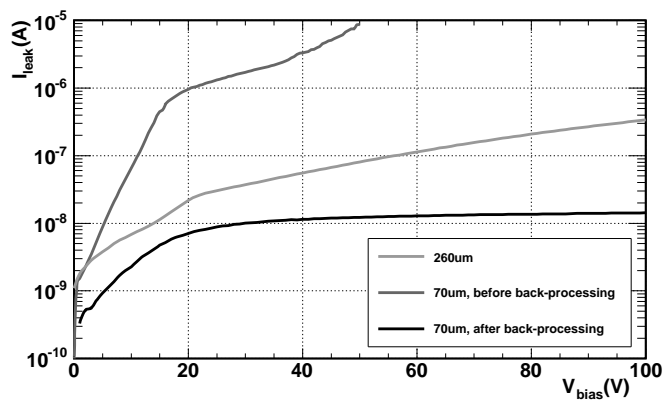,width=9.5cm}
\end{center}
\caption{I-V curve measured for the SOImager-2 before thinning (light grey line), after thinning and before back-side 
post-processing (dark grey line) and after $P$ implant and annealing (black line).}
\label{fig:IV}
\end{figure}

A set of $I-V$ measurements to determine the current flowing in the detector substrate as a function of the depletion 
voltage $V_d$ are performed by biasing the chip and monitoring the current with a DC source/monitor unit. In this measurement 
the depletion voltage is applied to the probe station plate. Two probes are used to measure the currents in the chip. 
The first measures the current to the pixel guard-ring grid and the other that to the $p+$ I/O implant, with the external 
guard-ring structure kept floating. Figure~\ref{fig:IV} compares the results obtained with a thick sensor with those 
of a thinned sensor before back-side post-processing and after $P$ implant and annealing. 
We observe a large increase in the leakage current after thinning, likely due to damage by the grinding process to the 
crystal structure. After back-plane implant and annealing the leakage current falls to values below those measured 
for un-thinned sensors.  

The measurement of the $C-V$ characteristics allows us to study the sensor depletion as a function of $V_d$.
For this measurement the pixel guard-ring grid is used as electrode and the detector area is assimilated to a single large 
diode. The guard-ring is kept at ground potential and the depletion voltage applied to the back-side. Since the determination 
of the effective area used to derive the capacitance is affected by a large uncertainty, we use the $C-V$ measurement only 
for establishing the voltage at which the sensor is fully depleted. From the evolution of the capacitance with the depletion 
voltage, we estimate that the detector is fully depleted for $V_d >$ 40~V. This is in agreement with what expected from the 
results of the 2010 beam test~\cite{beam2010-soi} and the resistivity deduced from the SRA analysis~\cite{xray-soi}. 

%We perform a 2D TCAD simulation of the device based on the doping profiles for the implants from the foundry and the SRA 
%profile for the back-plane contact. The electric field is computed across the substrate thickness for the central pixel in 
%a 10-pixel array for different $V_d$ values.

\subsection{Charge collection, detection efficiency and single point resolution}

\begin{figure}[hb!]
\begin{center}
\begin{tabular}{cc}
\epsfig{file=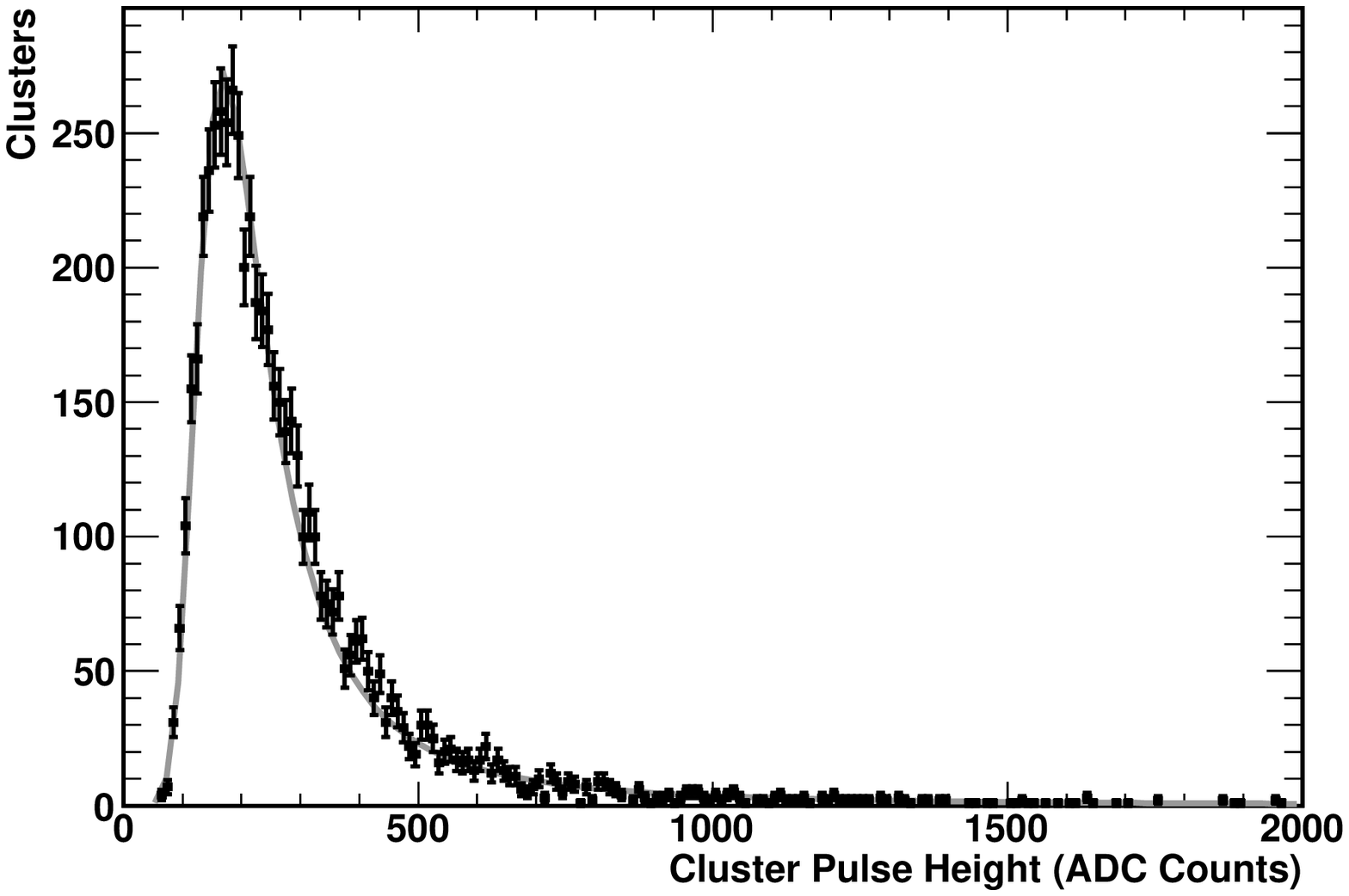,width=6.5cm} &
\epsfig{file=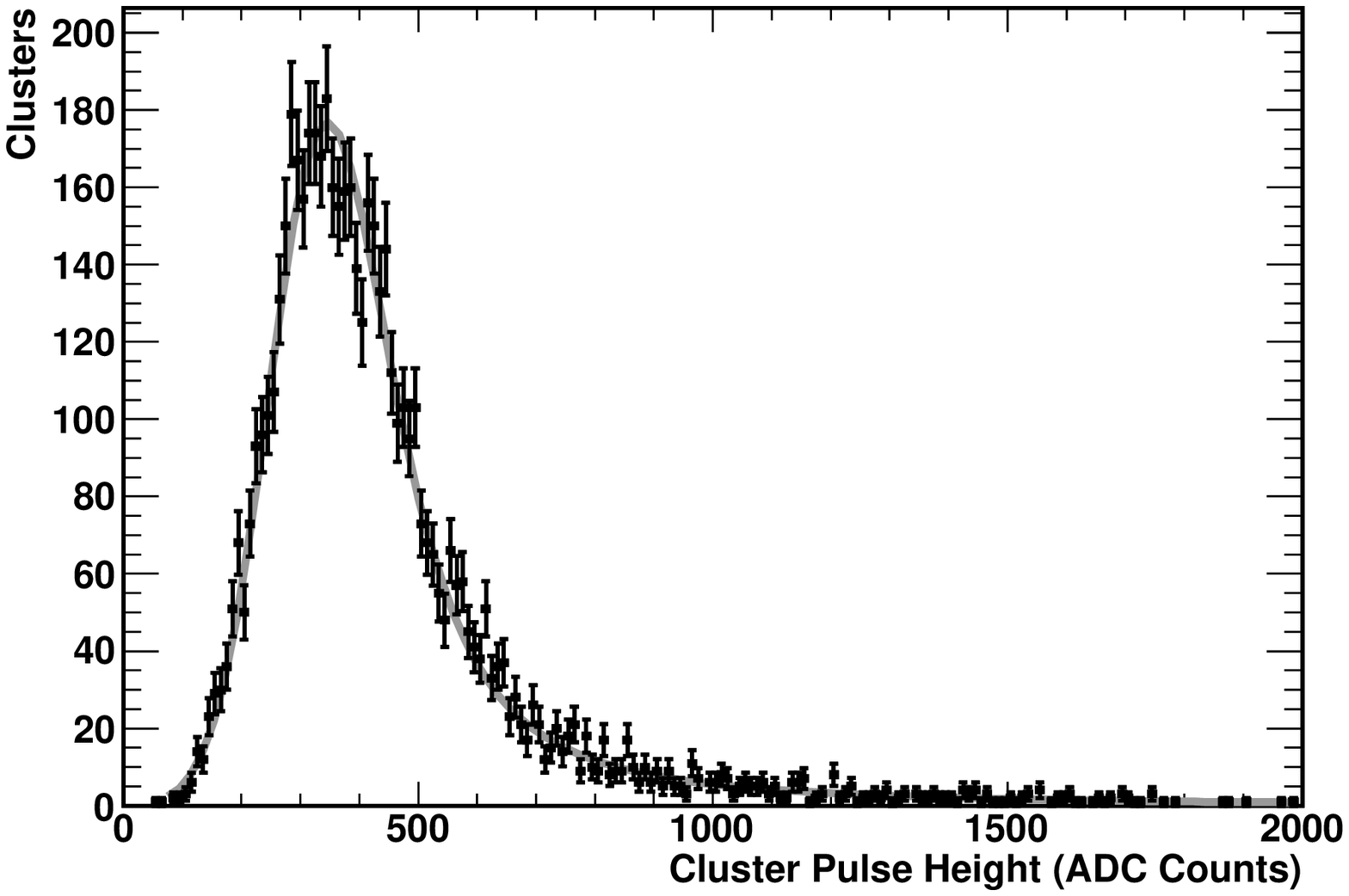,width=6.5cm} \\
\end{tabular}
\end{center}
\caption{Pulse height distribution of clusters reconstructed on the thinned (left) and thick (right) SOI sensor for 
cluster associated to reconstructed 300~GeV $\pi^-$ tracks for $V_d$ = 50~V. The curve represents the best fit of a Landau 
distribution folded with a Gaussian function.}
\label{fig:ph}
\end{figure}

The response of the thin sensor to minimum ionising particles is determined on signal clusters associated to 
reconstructed particle tracks having one hit on each of the three detector planes. The cluster pulse height of 
the thin SOI sensor is shown in the left panel of Figure~\ref{fig:ph}.
The most probable value of the charge deposited in the sensitive thickness of the sensor is (191$\pm$2)~ADC counts 
or (5980$\pm$100)~e$^-$.
This has to be compared to (314$\pm$3)~ADC counts or (9010$\pm$420)~e$^-$ obtained in a thick sensor for $V_d$ = 50~V, 
corresponding to a sensitive thickness of $\sim$100~$\mu$m~\cite{beam2010-soi} (see right panel of Figure~\ref{fig:ph}). 
The ratio of the pulse height values, 0.61$\pm$0.01, agrees well with the ratio of the estimated sensitive thickness of the two 
sensors of 0.62$\pm$0.05, where the quoted uncertainty includes the contribution from the thickness measurement of the thinned 
sensors and that on the depletion of the un-processed sensor. The measured noise of the thin sensor is (3.1$\pm$0.5)~ADC counts, 
consistent with that of the thick sensors of (3.3$\pm$0.3)~ADC counts. The observed average and most probable signal-to-noise 
ratio for clusters of minimum ionising particles is 28.2 and 23.7 respectively, to be compared to 47.4 and 43.0 measured for 
sensors of full thickness with $V_d$ = 50~V. The most probable value of the signal-to-noise ratio for the seed pixel 
is 26.2 and 45.1, respectively. This can be compared to 25.3$\pm$0.5 and 45.2$\pm$0.4 predicted by the simulation. Due to the 
decrease in signal pulse height and S/N caused by the thinner sensitive volume, compared to the thick sensors, we can expect a 
slight variation of the sensor efficiency and single point resolution.
The {\tt Geant-4} + {\tt PixelSim} simulation predicts a change of sensor efficiency from 0.998$^{+0.002}_{-0.012}$ for thick 
to 0.983$\pm$0.011 for the thin sensor at full depletion. We estimate the efficiency from the number of tracks reconstructed 
on the two layers of the doublet which have an associated hit on the thin sensor and find values from 0.94$\pm$0.03 to 
0.98$\pm$0.02 for $V_d$ = 50 to 90~V (see  Table~\ref{tab:res}), which agree with the simulation predictions. 
The single point resolution, $\sigma_{\mathrm{point}}$, is determined from the Gaussian width of the residual between 
the position of the reconstructed hit cluster and that of the particle track extrapolated from the doublet. 
\begin{figure}
\begin{center}
\begin{tabular}{cc}
\epsfig{file=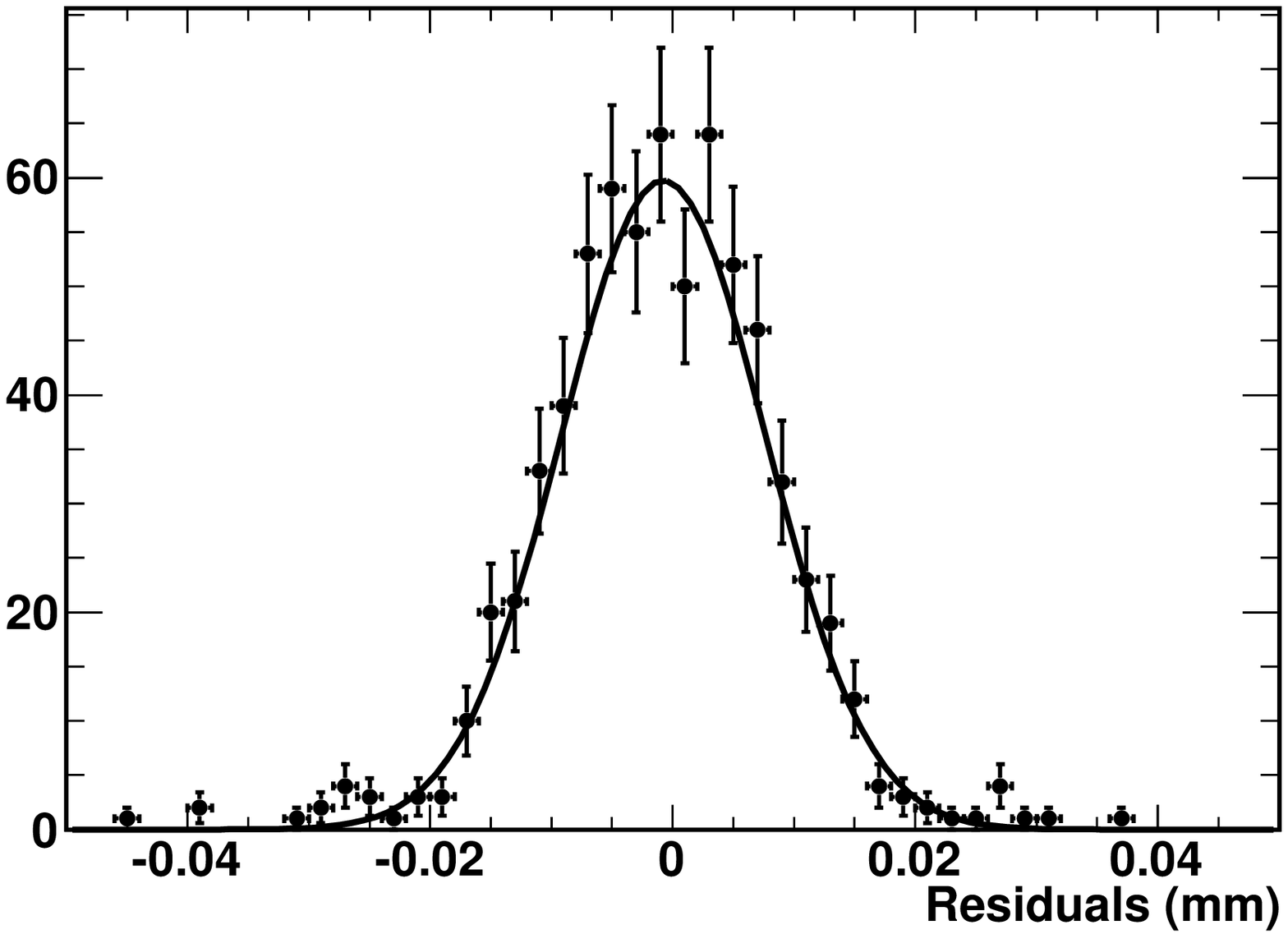,width=6.5cm} &
\epsfig{file=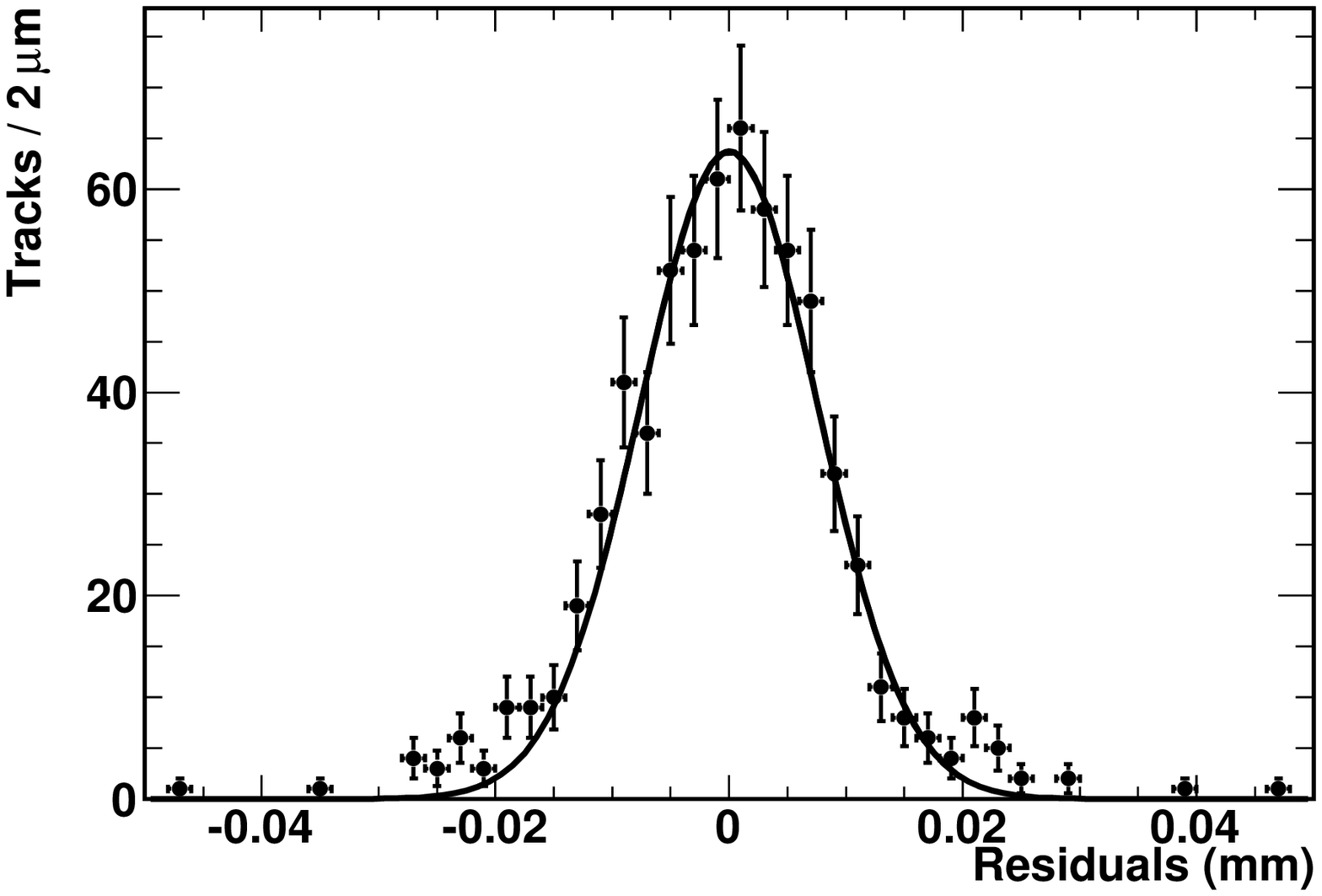,width=6.5cm} \\
\end{tabular}
\end{center}
\caption{Residual distribution for pion tracks reconstructed in the doublet and extrapolated upstream on the thin sensor, 
operated at (left) $V_d$ = 30~V and (right) 50~V. The fitted Gaussian curves have widths of (8.1$\pm$0.2)~$\mu$m and 
(7.6$\pm$0.3)~$\mu$m, respectively.}
\label{fig:res}
\end{figure}
The simulation predicts a single point resolution of (1.07$\pm$0.04)~$\mu$m for thick and (1.63$\pm$0.05)~$\mu$m for the 
fully depleted, thin sensor, where the change is due to the decrease in signal-to-noise ratio.
We measure residuals of (8.1$\pm$0.2)~$\mu$m and (7.6$\pm$0.3)~$\mu$m for $V_d$ = 30 V and 50~V, respectively (see Figure~\ref{fig:res}). 
By subtracting in quadrature the extrapolation resolution of the doublet, we obtain a single point resolution of the thin sensor 
of (1.7$\pm$0.5)~$\mu$m for $V_d$ = 50, where the uncertainty includes the statistical and the estimated systematic error 
from the extrapolation resolution.  These results are consistent with the simulation prediction. However, due to the long extrapolation 
to the thin sensor, the extrapolation resolution contribution is large and the single point resolution cannot be determined to much better 
than $\sim$0.5~$\mu$m. This accuracy is lower than that achieved in the analysis of the 2010 data (see ``Thick sensor'' results in 
Table~\ref{tab:res} and \cite{beam2010-soi}), where all the layers had the same resolution and both the residuals at the detector under 
test and the extrapolation resolution were sensitive to $\sigma_{\mathrm{point}}$. Still, results for $V_d \ge$ 50~V are consistent with 
a single point resolution of order of 1~$\mu$m and within the requirements for application to vertex tracking at future colliders. 
Results are summarised in Table~\ref{tab:res}.

\begin{table}
\caption{Measured average S/N ratio, detection efficiency and single point resolution $\sigma_{\mathrm{point}}$ for thin 
and thick SOI sensors.} 
\begin{center}
\begin{tabular}{|l|c|c|c|c|c|}
\hline
SOI    & $V_d$  & $d$         & Cluster            &  Efficiency   & $\sigma_{\mathrm{point}}$  \\
Sensor & (V)    & ($\mu$m)    &  $<\mathrm{S/N}>$  &               & ($\mu$m)     \\
\hline \hline
Thin    & 30    &  ~60$\pm$5  &   25.0             & 0.90$\pm$0.04 & 3.1~$\pm$0.80  \\
        & 50    &  ~64$\pm$3  &   28.2             & 0.94$\pm$0.03 & 1.7~$\pm$0.50  \\ 
        & 70    &  ~64$\pm$3  &   28.8             & 0.96$\pm$0.03 & 1.8~$\pm$0.60  \\ 
        & 90    &  ~64$\pm$3  &   31.2             & 0.98$\pm$0.02 & 1.9~$\pm$0.70  \\ \hline
Thick   & 30    &  ~60$\pm$8  &   23.3             & 0.89$\pm$0.03 & 1.36$\pm$0.04  \\
        & 50    &  103$\pm$5  &   47.4             & 0.98$^{+0.02}_{-0.04}$ & 1.12$\pm$0.03  \\
        & 70    &  122$\pm$5  &   52.7             & 0.99$^{+0.01}_{-0.05}$ & 1.07$\pm$0.05  \\
\hline
\end{tabular}
\end{center}
\label{tab:res}
\end{table}

\section{Conclusions}

A pixel sensor in SOI technology on high-resistivity substrate thinned to 70~$\mu$m with a thin phosphor 
layer contact implanted on the back-plane has been characterised with high momentum charged pions at the CERN 
SPS. The sensor is operated in full depletion. Its response is compared to that of un-processed sensors of the 
same design. The measured cluster pulse height corresponds to (5980$\pm$100)~e$^-$ and the ratio of the pulse 
height measured in this detector to that in a thick sensor is found to be 0.61$\pm$0.01, which is consistent 
with the ratio of the estimated depleted thicknesses in the two sensors of 0.62$\pm$0.05. The measured average 
cluster signal-to-noise ratio for the thin sensor is $\sim$30 for $V_d \ge$ 50~V. The detection efficiency is 
determined to be 0.96$\pm$0.02 and the point resolution (1.8$\pm$0.4)~$\mu$m for clusters associated to reconstructed 
particle tracks, at  $V_d \ge$ 50~V, in agreement with simulation predictions. These results show that a thin, fully 
depleted SOI pixel provides charged particle detection capability with large signal-to-noise ratio and detection 
efficiency and achieves a single point resolution of order of 1~$\mu$m. 

\section*{Acknowledgements}

This work was supported by the director, Office of Science, of the 
U.S. Department of Energy under Contract No.DE-AC02-05CH11231 and 
by INFN, Italy. We are grateful to Y.~Arai for his effective 
collaboration and support in the SOIPIX activities. We are indebted 
to I.~Efthymiopoulos and  M.~Jeckel for support on the SPS beam-line and 
to A.~Behrens and E.~Lacroix for performing the detector system alignment.  
We also thank the CERN IT department for computing support.

\end{document}